\documentclass[a4paper,english]{article}
\usepackage[T1]{fontenc}
\usepackage[latin1]{inputenc}
\usepackage{float}
\usepackage{graphicx}

\makeatletter

 \newenvironment{lyxlist}[1]
   {\begin{list}{}
     {\settowidth{\labelwidth}{#1}
      \setlength{\leftmargin}{\labelwidth}
      \addtolength{\leftmargin}{\labelsep}
      }}
   {\end{list}}

\renewcommand\theparagraph    {
\@arabic\c@paragraph}

\usepackage{babel}
\makeatother
\begin{document}

\title{Does special relativity theory tell us anything new about space and
time?}

\author{László E. Szabó\\
 \emph{}\emph{\small Theoretical Physics Research Group of the Hungarian
Academy of Sciences} {\small }\\
 {\small }\emph{\small Department of History and Philosophy of Science}
{\small }\\
 {\small }\emph{\small Eötvös University, Budapest}{\small }\\
 {\small }\emph{\small E-mail: leszabo@hps.elte.hu}}

\date{~}

\maketitle
\begin{abstract}
It will be shown that, in comparison with the pre-relativistic Galileo-invariant
conceptions, special relativity tells us nothing new about the geometry
of space-time. It simply calls something else {}``space-time'',
and this something else has different properties. All statements of
special relativity about those features of reality that correspond
to the original meaning of the terms {}``space'' and {}``time''
are identical with the corresponding traditional pre-relativistic
statements. It will be also argued that special relativity and Lorentz
theory are completely identical in both senses, as theories about
space-time and as theories about the behaviour of moving physical
objects.

\medskip{}
\end{abstract}

\newpage
\section*{Prolog}

Consider the following  definitions of electrodynamical quantities:%
\begin{figure}[h]
\begin{center}\includegraphics[%
  width=0.40\columnwidth]{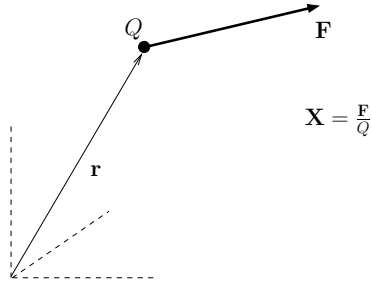}\end{center}

\caption{$\mathbf{X}$ is defined as the force felt by the unit test charge\label{cap:E}}
\end{figure}

\begin{lyxlist}{00.00.0000}
\item [$\mathbf{X}\left(\mathbf{r}\right)$]Locate a test charge $Q$ at
point $\mathbf{r}$ and measure the force $\mathbf{F}$ felt by the
charge. $\mathbf{X}\left(\mathbf{r}\right)=\frac{\mathbf{F}}{Q}$
(Fig~\ref{cap:E}).
\item [$\mathbf{Y}\left(\mathbf{r}\right)$]Locate two contacting metal
plates of area $A$ at point $\mathbf{r}$. Separate them and measure
the influence charge $Q$ on one of the plates. $Y(\mathbf{r})=\frac{Q}{A}$.
The direction of $\mathbf{Y}(\mathbf{r})$ is determined by the normal
vector of the plates, when the charge separation is maximal (Fig~\ref{cap:D}).%
\begin{figure}[h]
\begin{center}\includegraphics[%
  width=0.50\columnwidth]{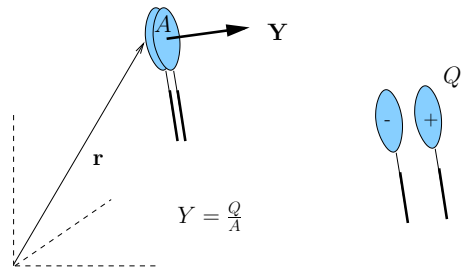}\end{center}

\caption{$\mathbf{Y}$ is defined by means of the influence charge divided
by the surface\label{cap:D}}
\end{figure}

\end{lyxlist}
It is a well known empirical fact that these quantities are not independent
of each other. For the sake of simplicity, assume the simplest material
equation \begin{equation}
\mathbf{Y}=\varepsilon\mathbf{X}\label{eq:anyagi}\end{equation}
where $\varepsilon$, called dielectric constant, is a scalar field
characterising the medium. 

Traditionally, in phenomenological electrodynamics, physical quantity
$\mathbf{X}$ is called `electric field strength' and denoted by $\mathbf{E}$,
and $\mathbf{Y}$ is called `electric displacement' and denoted by
$\mathbf{D}$. Due to the material equation (\ref{eq:anyagi}) one
can eliminate one of the field variables. 

\begin{figure}[H]
\begin{center}\includegraphics[%
  width=0.40\columnwidth]{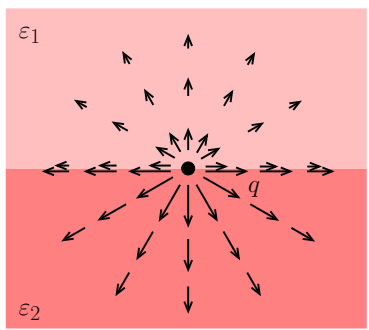}\end{center}

\caption{The `electric field strength' of the static electric field around
a point charge $q$ located at the border of two materials of dielectric
constants $\varepsilon_{1}$ and $\varepsilon_{2}$\label{cap:kettos1}}
\end{figure}

Imagine a text book (I shall refer to it as the {}``old'' one),
which only uses $\mathbf{E}$. The equations of electrostatics are
written as follows:\begin{eqnarray}
\textrm{div }\mathbf{\varepsilon E} & = & \rho\label{eq:tradMax1}\\
\textrm{rot }\mathbf{E} & = & 0\label{eq:tradMax2}\end{eqnarray}
For example, the book contains the following exercise and solution:

\begin{quote}
\textbf{Exercise}~~~Consider the static electric field around a
point charge $q$ located at the border of two materials of dielectric
constant $\varepsilon_{1}$ and $\varepsilon_{2}$. Is the electric
field strength spherically symmetric, or not?

\textbf{Solution} %
\begin{figure}[b]
\begin{center}\includegraphics[%
  width=0.40\columnwidth]{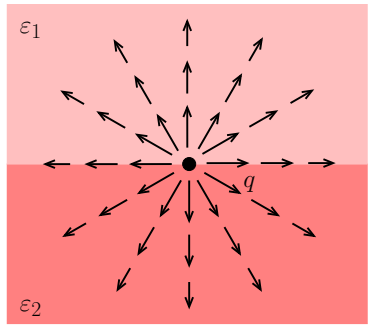}\end{center}

\caption{The `electric field strength' of the static electric field around
a point charge $q$ located at the border of two materials of dielectric
constants $\varepsilon_{1}$ and $\varepsilon_{2}$ \label{cap:kettos2}}
\end{figure}
 (see Fig~\ref{cap:kettos2})\begin{eqnarray}
\mathbf{E}_{1} & = & \frac{1}{2\pi\left(\varepsilon_{1}+\varepsilon_{2}\right)}\frac{q}{r^{3}}\mathbf{r}\label{eq:tradmegoldas1}\\
\mathbf{E}_{2} & = & \frac{1}{2\pi\left(\varepsilon_{1}+\varepsilon_{2}\right)}\frac{q}{r^{3}}\mathbf{r}\label{eq:tradmegoldas2}\end{eqnarray}
Consequently,\begin{equation}
\textrm{The electric field strength is spherically symmetric.}\label{eq:szimmetrikus}\end{equation}

\end{quote}
Now, imagine a new electrodynamics text book which is non-traditional
in the following sense: it uses only field variable $\mathbf{Y}$
(traditionally called `electric displacement' and denoted by $\mathbf{D}$),
but it systematically calls $\mathbf{Y}$ `electric field strength'
and denotes it by $\mathbf{E}$. Accordingly, the equations of electrostatics
are written as follows:\begin{eqnarray}
\textrm{div }\mathbf{E} & = & \rho\label{eq:Max1}\\
\textrm{rot }\mathbf{\frac{E}{\varepsilon}} & = & 0\label{eq:Max2}\end{eqnarray}
This new book also contains the above exercise, but with the following
solution:

\begin{quote}
\textbf{Solution} (see Fig~\ref{cap:kettos1})

\begin{eqnarray}
\mathbf{E}_{1} & = & \frac{\varepsilon_{1}}{2\pi\left(\varepsilon_{1}+\varepsilon_{2}\right)}\frac{q}{r^{3}}\mathbf{r}\label{eq:megoldas1}\\
\mathbf{E}_{2} & = & \frac{\varepsilon_{2}}{2\pi\left(\varepsilon_{1}+\varepsilon_{2}\right)}\frac{q}{r^{3}}\mathbf{r}\label{eq:megoldas2}\end{eqnarray}
Consequently,\begin{equation}
\textrm{The electric field strength is not spherically symmetric.}\label{eq:nem szimmetrikus}\end{equation}

\end{quote}
Now, does sentence (\ref{eq:nem szimmetrikus}) of the new book contradict
to sentence (\ref{eq:szimmetrikus}) of the old book? Is it true that
the theory described in the new book is a \emph{new theory} of electromagnetism?
Of course, not. Seemingly the two sentences contradict to each other,
on the level of the words. However, in order to clarify the meaning
of sentence (\ref{eq:nem szimmetrikus}) and (\ref{eq:szimmetrikus}),
one has to go back to the first pages of the corresponding book and
clarify the definition of the physical quantity called `electric field
strength'. And it will be clear that the term `electric field strength'
stands for two different physical quantities in the two books. Moreover,
both text books provide complete descriptions of electromagnetic phenomena.
Therefore, although the theory in the old book does not use the field
variable $\mathbf{Y}$, it is capable to account for the physical
phenomena by which physical quantity $\mathbf{Y}$ is empirically
defined. It is capable to determine the influence charge on the separated
plates (by calculating $\varepsilon EA$). In other words, it is capable
to determine the value of $\mathbf{Y}$, that is, the value of what
the new book calls `electric field strength'. And vice versa, on the
basis of the theory described in the new book one can calculate the
force felt by a unit test charge (by calculating $\frac{\mathbf{E}}{\varepsilon}$),
that is, one can predict the value of $\mathbf{X}$, what the old
book calls `electric field strength'. And both, the theory in the
old book and the theory in the new book have the same predictions
for both, $\mathbf{X}$ and $\mathbf{Y}$. That is to say, although
they use different terminology, the two text books contain the same
electrodynamics, they provide the same description of physical reality.

\newpage
\section{Introduction}

\medskip{}

\begin{flushright}\begin{minipage}[c]{0.90\columnwidth}\noindent \emph{\small I have for long thought that if I had the opportunity
to teach this subject, I would emphasize the continuity with earlier
ideas. Usually it is the discontinuity which is stressed, the radical
break with more primitive notions of space and time. Often the result
is to destroy completely the confidence of the student in perfectly
sound and useful concepts already acquired.} {\small (From J. S. Bell:
{}``How to teach special relativity}\emph{\small '',} {\small Bell
1987, p. 67.)}\end{minipage}\end{flushright}

\medskip{}

It is widely believed that the principal difference between Einstein's
special relativity and its contemporary rival Lorentz theory was that
while the Lorentz theory%
\footnote{I use the term {}``Lorentz theory'' as classification to refer to
the similar approaches of Lorentz, FitzGerald, and Poincaré, that
save the classical Galilei covariant conceptions of space and time
by explaining the null result of the Michelson--Morley experiment
and other similar experimental findings through the physical distortions
of moving objects (first of all of moving measuring-rods and clocks),
no matter whether these physical distortions are simply hypothesised
in the theory, or prescribed by some {}``principle'' like Lorentz's
principle, or they are constructively derived from the behaviour of
the molecular forces. From the point of view of my recent concerns
what is important is the logical possibility of such an alternative
theory. Although, Lorentz's 1904 paper is very close to be a good
historic example.%
} was also capable of {}``explaining away'' the null result of the
Michelson--Morley experiment and other experimental findings by means
of the distortions of moving measuring-rods and moving clocks, special
relativity revealed more fundamental new facts about the geometry
of space-time behind these phenomena. According to this widespread
view, special relativity theory has radically changed our conceptions
about space and time by claiming that space-time is not like an $E^{3}\times E^{1}$
space, as was believed in classical physics, but it is a four dimensional
Minkowski space $M^{4}$. One can express this revolutionary change
by the following logical schema: Earlier we believed in $G_{1}\left(\mathcal{M}\right)$,
where $\mathcal{M}$ stands for space-time and $G_{1}$ denotes some
predicate (like $E^{3}\times E^{1}$). Then we discovered that $\neg G_{1}\left(\mathcal{M}\right)$
but $G_{2}\left(\mathcal{M}\right)$, where $G_{2}$ denotes a predicate
different from $G_{1}$ (something like $M^{4}$). 

Contrary to this common view, the first main thesis in this paper
is the following:

\medskip{}
\noindent \textbf{Thesis~1.}\emph{~~~In comparison with the pre-relativistic
Galileo-invariant conceptions, special relativity tells us nothing
new about the geometry of space-time. It simply calls something else
{}``space-time'', and this something else has different properties.
All statements of special relativity about those features of reality
that correspond to the original meaning of the terms {}``space''
and {}``time'' are identical with the corresponding traditional
pre-relativistic statements. }
\medskip{}

\noindent Thus the only new factor in the special relativistic account
of space-time is the decision to designate something else {}``space-time''.
In other words: Earlier we believed in $G_{1}\left(\mathcal{M}\right)$.
Then we discovered for some $\widetilde{\mathcal{M}}\neq\mathcal{M}$
that $\neg G_{1}\left(\widetilde{\mathcal{M}}\right)$ but $G_{2}\left(\widetilde{\mathcal{M}}\right)$.
Consequently, it still holds that $G_{1}\left(\mathcal{M}\right)$. 

So the real novelty in special relativity is some $G_{2}\left(\widetilde{\mathcal{M}}\right)$.
As we will see, this is nothing but the description of the physical
behaviour of moving measuring-rods and clocks. It will be also argued,
however, that $G_{2}\left(\widetilde{\mathcal{M}}\right)$ is exactly
what Lorentz theory claims. More exactly, as my second main thesis
asserts, both theories claim that $G_{1}\left(\mathcal{M}\right)\& G_{2}\left(\widetilde{\mathcal{M}}\right)$:

\medskip{}
\noindent \textbf{Thesis~2.}\emph{~~~Special relativity and Lorentz
theory are completely identical in both senses, as theories about
space-time and as theories about the behaviour of moving physical
objects.}
\medskip{}

\section{On the meaning of the question {}``What is space-time like?''}

A theory \emph{about} space-time describes a certain group of objective
features of physical reality, which we call (the structure of) space-time.
According to classical physics, space-time can be described through
a geometrical structure like $E^{3}\times E^{1}$, where $E^{3}$
is a three-dimensional Euclidean space for space, and $E^{1}$ is
a one-dimensional Euclidean space for time, with two independent invariant
metrics corresponding to the space and time intervals. In contrast,
special relativity claims that space-time---understood as the same
objective features of physical reality---is something different: it
can be described through a Minkowski geometry (such that the simultaneity
is expressed via orthogonality with respect of the 4-metric of the
Minkowski space, etc.) 

Physics describes objective features of reality by means of physical
quantities. Our scrutiny will therefore start by clarifying how classical
physics and relativity theory define the space and time tags assigned
to an arbitrary event. It will be seen that these empirical definitions
are different. 

The empirical definition of a physical quantity requires an \emph{etalon}
measuring equipment and a precise description of the operation how
the quantity to be defined is measured. For example, assume we choose,
as the \emph{etalon} measuring-rod, the meter stick that is lying
in the International Bureau of Weights and Measures (BIPM) in Paris.
Also assume---this is another convention---that {}``time'' is defined
as a physical quantity measured by the standard clock also sitting
in the BIPM. When I use the word {}``convention'' here, I mean the
semantical freedom we have in the use of the uncommitted signs {}``distance''
and {}``time''---a freedom what Grünbaum (1974, p. 27) calls {}``trivial
semantical conventionalism''. 

Now we are going to describe the operations by which we define the
space and time tags of an arbitrary event $A$, relative to the reference
frame $K$ in which the the \emph{etalons} are at rest, and to another
reference fame $K'$ which is moving (at constant velocity $v$) relative
to $K$. For the sake of simplicity consider only one space dimension
and assume that the origin of both $K$ and $K'$ is at the BIPM at
the initial moment of time.

\paragraph*{(D1)~Time tag in $K$ according to classical physics}

\begin{quote}
Take a synchronised copy of the standard clock at rest in the BIPM,
and slowly%
\footnote{{}``Slowly'' means that we move the clock from one place to the
other over a long period of time, according to the reading of the
clock itself. The reason is to avoid the loss of phase accumulated
by the clock during its journey.%
} move it to the locus of event $A$. The time tag $t^{K}\left(A\right)$
is the reading of the transfered clock when $A$ occurs.%
\footnote{With this definition we actually use the standard {}``$\varepsilon=\frac{1}{2}$-synchronisation''.
I do not want to enter now into the question of the conventionality
of simultaneity, which is a hotly debated problem, in itself. (See
Reichenbach 1956; Grünbaum 1974; Salmon 1977; Malament 1977; Friedman
1983.)%
}
\end{quote}

\paragraph*{(D2)~Space tag in $K$ according to classical physics}

\begin{quote}
The space tag $x^{K}(A)$ of event $A$ is is the distance from the
origin of $K$ of the locus of $A$ along the $x$-axis%
\footnote{The straight line is defined by a light beam.%
} measured by superposing the standard measuring-rod, being always
at rest relative to $K$. 
\end{quote}

\paragraph*{(D3)~Time tag in $K$ according to special relativity}

\begin{quote}
Take a synchronised copy of the standard clock at rest in the BIPM,
and slowly move it to the locus of event $A$. The time tag $\widetilde{t}^{K}\left(A\right)$
is the reading of the transfered clock when $A$ occurs. 
\end{quote}

\paragraph*{(D4)~Space tag in $K$ according to special relativity}

\begin{quote}
The space tag $\widetilde{x}^{K}(A)$ of event $A$ is the distance
from the origin of $K$ of the locus of $A$ along the $x$-axis measured
by superposing the standard measuring-rod, being always at rest relative
to $K$. 
\end{quote}

\paragraph*{(D5)~Space and time tags of an event in $K'$ according to classical
physics}

\begin{quote}
The space tag of event $A$ relative to the frame $K'$ is $x^{K'}(A):=x^{K}(A)-vt^{K}(A)$,
where $v=v^{K}(K')$ is the velocity of $K'$ relative to $K$ in
the sense of definition (D8).\\
 The time tag of event $A$ relative to the frame $K'$ is $t^{K'}(A):=t^{K}(A)$
\end{quote}

\paragraph*{(D6)~Time tag in $K'$ according to special relativity}

\begin{quote}
Take a synchronised copy of the standard clock at rest in the BIPM,
gently accelerate it from $K$ to $K'$ and set it to show $0$ when
the origins of $K$ and $K'$ coincide. Then slowly (relative to $K'$)
move it to the locus of event $A$. The time tag $\widetilde{t}^{K'}\left(A\right)$
is the reading of the transfered clock when $A$ occurs. 
\end{quote}

\paragraph*{(D7)~Space tag in $K'$ according to special relativity}

\begin{quote}
The space tag $\widetilde{x}^{K'}(A)$ of event $A$ is the distance
from the origin of $K'$ of the locus of $A$ along the $x$-axis
measured by superposing the standard measuring-rod, being always at
rest relative to $K'$, in just the same way as if all were at rest. 
\end{quote}

\paragraph*{(D8)~Velocities in the different cases}

\begin{quote}
Velocity is a quantity derived from the above defined space and time
tags:\begin{eqnarray*}
v^{K} & = & \frac{\Delta x^{K}}{\Delta t^{K}}\\
\widetilde{v}^{K} & = & \frac{\Delta\widetilde{x}^{K}}{\Delta\widetilde{t}^{K}}\\
v^{K'} & = & \frac{\Delta x^{K'}}{\Delta t^{K'}}\\
\widetilde{v}^{K'} & = & \frac{\Delta\widetilde{x}^{K'}}{\Delta\widetilde{t}^{K'}}\end{eqnarray*}

\end{quote}
With these empirical definitions, in every inertial frame we define
four different quantities for each event, such that:\begin{eqnarray}
x^{K}(A) & \equiv & \widetilde{x}^{K}(A)\label{eq:identity1}\\
t^{K}(A) & \equiv & \widetilde{t}^{K}(A)\label{eq:identity2}\\
x^{K'}(A) & \not\equiv & \widetilde{x}^{K'}(A)\label{eq:op-ineq1}\\
t^{K'}(A) & \not\equiv & \widetilde{t}^{K'}(A)\label{eq:op-ineq2}\end{eqnarray}
 where $\equiv$ denotes the identical operational definition.

In spite of the different operational definitions, it could be a \emph{contingent}
fact of nature that $x^{K'}(A)=\widetilde{x}^{K'}(A)$ and/or $t^{K'}(A)=\widetilde{t}^{K'}(A)$
for every event $A$.%
\footnote{Let me illustrate this with an example. The inertial mass $m_{i}$
and gravitational mass $m_{g}$ are two quantities having different
experimental definitions. But, it is a contingent fact of nature (experimentally
proved by Eötvös around 1900) that, for any object, the two masses
are equal, $m_{i}=m_{g}$.%
} But a little reflection reveals that this is not the case. It follows
from special relativity that $\widetilde{x}^{K}(A),\widetilde{t}^{K}(A)$
are related with $\widetilde{x}^{K'}(A),\widetilde{t}^{K'}(A)$ through
the Lorentz transformation, while $x^{K}(A),t^{K}(A)$ are related
with $x^{K'}(A),t^{K'}(A)$ through the corresponding Galilean transformation,
therefore, taking into account identities (\ref{eq:identity1})--(\ref{eq:identity2}),
$x^{K'}(A)\neq\widetilde{x}^{K'}(A)$ and $t^{K'}(A)\neq\widetilde{t}^{K'}(A)$,
if $v\neq0$. 

Thus, our first partial conclusion is that \emph{different physical
quantities are called {}``space'' tag, and similarly, different
physical quantities are called {}``time'' tag in special relativity}
\emph{and in classical physics}.%
\footnote{This was first recognised by Bridgeman (1927, p. 12), although he
did not investigate the further consequences of this fact.%
} In order to avoid further confusion, from now on space and time tags
will mean the physical quantities defined in (D1), (D2) and (D5)---according
to the usage of the terms in classical physics---, and {}``space''
and {}``time'' in the sense of the relativistic definitions (D3),
(D4), (D6) and (D7) will be called $\widetilde{\textrm{space}}$ and
$\widetilde{\textrm{time}}$.

Special relativity theory makes \emph{different} assertions about
somethings which are \emph{different} from space and time. In our
symbolic notation, classical physics claims $G_{1}\left(\mathcal{M}\right)$
about $\mathcal{M}$ and relativity theory claims $G_{2}\left(\widetilde{\mathcal{M}}\right)$
about some other features of reality $\widetilde{\mathcal{M}}$. The
question is what special relativity and classical physics say when
they are making assertions about the same things.

\section{Special relativity does not tell us anything new about space and
time}

Classical physics and relativity theory would be different theories
of space and time if they accounted for physical quantities $x$ and
$t$ differently. If there were any event $A$ and any inertial frame
of reference $K^{\star}$ in which the space or time tag assigned
to the event by special relativity, $\left[x^{K^{\star}}(A)\right]_{relativity}$,
$\left[t^{K^{\star}}(A)\right]_{relativity}$, were different from
the similar tags assigned by classical physics, $\left[x^{K^{\star}}(A)\right]_{classical}$,
$\left[t^{K^{\star}}(A)\right]_{classical}$. If, for example, there
were any two events simultaneous in relativity theory which were not
simultaneous according to classical physics, or vice versa---to touch
on a sore point. But a little reflection shows that this is not the
case. Taking into account operational identities (\ref{eq:identity1})--(\ref{eq:identity2}),
one can calculate the relativity theoretic prediction for the outcome
of operations described in (D1), (D2) and (D5), that is, the relativity
theoretic prediction for $x^{K'}(A)$:\begin{equation}
\left[x^{K'}(A)\right]_{relativity}=\widetilde{x}^{K}(A)-\widetilde{v}^{K}(K')\widetilde{t}^{K}(A)\label{eq:ugyanaz0}\end{equation}
 the value of which is equal to \begin{eqnarray}
 &  & x^{K}(A)-v^{K}(K')t^{K}(A)=\left[x^{K'}(A)\right]_{classical}\label{eq:ugyanaz1}\end{eqnarray}
 Similarly, \begin{equation}
\left[t^{K'}(A)\right]_{relativity}=\widetilde{t}^{K}(A)=t^{K}(A)=\left[t^{K'}(A)\right]_{classical}\label{eq:ugyanaz2}\end{equation}
This completes the proof of Thesis~1.

\section{Lorentz theory and special relativity are completely identical theories\label{sec:Lorentz-theory-and}}

Since Lorentz theory adopts the classical theory of space-time, it
does not differ from special relativity in its assertions about space
and time. However, beyond what special relativity claims about space
and time, it also has another claim---$G_{2}\left(\widetilde{\mathcal{M}}\right)$---about
$\widetilde{\textrm{space}}$ and $\widetilde{\textrm{time}}$. In
order to prove what Thesis~2 asserts, that is to say the complete
identity of Lorentz theory and of special relativity, we also have
to show that the two theories have identical assertions about $\widetilde{x}$
and $\widetilde{t}$, that is,\begin{eqnarray*}
\left[\widetilde{x}^{K'}(A)\right]_{relativity} & = & \left[\widetilde{x}^{K'}(A)\right]_{LT}\\
\left[\widetilde{t}^{K'}(A)\right]_{relativity} & = & \left[\widetilde{t}^{K'}(A)\right]_{LT}\end{eqnarray*}
 According to relativity theory, the $\widetilde{\textrm{space}}$
and $\widetilde{\textrm{time}}$ tags in $K'$ and in $K$ are related
through the Lorentz transformations. From (\ref{eq:identity1})--(\ref{eq:identity2})
one can deduce:\begin{eqnarray}
\left[\widetilde{t}^{K'}(A)\right]_{relativity} & = & \frac{t^{K}(A)-\frac{v\, x^{K}(A)}{c^{2}}}{\sqrt{1-\frac{v^{2}}{c^{2}}}}\label{eq:Lorentz1}\\
\left[\widetilde{x}^{K'}(A)\right]_{relativity} & = & \frac{x^{K}(A)-v\, t^{K}(A)}{\sqrt{1-\frac{v^{2}}{c^{2}}}}\label{eq:Lorentz2}\end{eqnarray}

On the other hand, taking the assumptions of Lorentz theory that the
standard clock slows down by factor $\sqrt{1-\frac{v^{2}}{c^{2}}}$
and that a rigid rod suffers a contraction by factor $\sqrt{1-\frac{v^{2}}{c^{2}}}$
when they are gently accelerated from $K$ to $K'$, one can \emph{}directly
calculate the $\widetilde{\textrm{space}}$ tag $\widetilde{x}^{K'}(A)$
and the $\widetilde{\textrm{time}}$ tag $\widetilde{t}^{K'}(A)$,
following the descriptions of operations in (D6) and (D7). 

First, let us calculate the reading of the clock slowly transported
in $K'$ from the origin to the locus of an event $A$. The clock
is moving with a varying velocity%
\footnote{For the sake of simplicity we continue to restrict our calculation
to one space dimension. For the general calculation of the phase shift
suffered by moving clocks, see Jánossy 1971, pp. 142--147.%
}\[
v_{C}^{K}(t^{K})=v+w^{K}(t^{K})\]
 where $w^{K}(t^{K})$ is the velocity of the clock relative to $K'$,
that is, $w^{K}(0)=0$ when it starts at $x_{C}^{K}(0)=0$ (as we
assumed, $t^{K}=0$ and the transported clock shows $0$ when the
origins of $K$ and $K'$ coincide) and $w^{K}(t_{1}^{K})=0$ when
the clock arrives at the place of $A$. The reading of the clock at
the time $t_{1}^{K}$ will be\begin{equation}
T=\int_{0}^{t_{1}^{K}}\sqrt{1-\frac{\left(v+w^{K}(t)\right)^{2}}{c^{2}}}\, dt\label{eq:integral}\end{equation}
 Since $w^{K}$ is small we may develop in powers of $w^{K}$, and
we find from (\ref{eq:integral}) when neglecting terms of second
and higher order\begin{equation}
T=\frac{t_{1}^{K}-\frac{\left(t_{1}^{K}v+\int_{0}^{t_{1}^{K}}w^{K}(t)\, dt\right)v}{c^{2}}}{\sqrt{1-\frac{v^{2}}{c^{2}}}}=\frac{t^{K}(A)-\frac{x^{K}(A)v}{c^{2}}}{\sqrt{1-\frac{v^{2}}{c^{2}}}}\label{eq:readingoftranfered}\end{equation}
 (where, without loss of generality, we take $t_{1}^{K}=t^{K}(A)$).
Thus, according to the definition of $\widetilde{t}$, we have\[
\left[\widetilde{t}^{K'}(A)\right]_{LT}=\frac{t^{K}(A)-\frac{v\, x^{K}(A)}{c^{2}}}{\sqrt{1-\frac{v^{2}}{c^{2}}}}\]
which is equal to $\left[\widetilde{t}^{K'}(A)\right]_{relativity}$
in (\ref{eq:Lorentz1}).

Now, taking into account that the length of the co-moving meter stick
is only $\sqrt{1-\frac{v^{2}}{c^{2}}}$, the distance of event $A$
from the origin of $K$ is the following:\[
x^{K}(A)=t^{K}(A)v+\widetilde{x}^{K'}(A)\sqrt{1-\frac{v^{2}}{c^{2}}}\]
 and thus\[
\left[\widetilde{x}^{K'}(A)\right]_{LT}=\frac{x^{K}(A)-v\, t^{K}(A)}{\sqrt{1-\frac{v^{2}}{c^{2}}}}=\left[\widetilde{x}^{K'}(A)\right]_{relativity}\]
This completes the proof. The two theories make completely identical
assertions not only about the space and time tags $x,t$ but also
about the $\widetilde{\textrm{space}}$ and $\widetilde{\textrm{time}}$
tags $\tilde{x},\tilde{t}$. 

Consequently, there is full agreement between the Lorentz theory and
special relativity theory in the following statements:

\begin{lyxlist}{0.0.0}
\item [(a)]$\widetilde{\textrm{Velocity}}$---which is called {}``velocity''
by relativity theory---is not an additive quantity,\[
\widetilde{v}^{K'}(K''')=\frac{\widetilde{v}^{K'}(K'')+\widetilde{v}^{K''}(K''')}{1+\frac{\widetilde{v}^{K'}(K'')\widetilde{v}^{K''}(K''')}{c^{2}}}\]
while velocity---that is, what we traditionally call {}``velocity''---is
an additive quantity,\[
v^{K'}(K''')=v^{K'}(K'')+v^{K''}(K''')\]
 where $K',K'',K'''$ are arbitrary three frames. For example, \[
v^{K'}(light\, signal)=v^{K'}(K'')+v^{K''}(light\, signal)\]

\item [(b)]The $\left(\widetilde{x}_{1},\widetilde{x}_{2},\widetilde{x}_{3},\widetilde{t}\right)$-map
of the world can be conveniently described through a Minkowski geometry,
such that the {}``$\widetilde{t}$-simultaneity'' can be described
through the orthogonality with respect to the 4-metric of the Minkowski
space, etc.
\item [(c)]The $\left(x_{1},x_{2},x_{3},t\right)$-map of the world, can
be conveniently described through a traditional space-time geometry
like $E^{3}\times E^{1}$.
\item [(d)]The velocity of light is not the same in all inertial frames
of reference.
\item [(e)]The $\widetilde{\textrm{velocity}}$ of light is the same in
all inertial frames of reference.
\item [(f)]Time and distance are invariant, the reference frame independent
concepts, $\widetilde{\textrm{time}}$ and $\widetilde{\textrm{distance}}$
are not.
\item [(g)]$t$-simultaneity is an invariant, frame-independent concept,
while $\widetilde{t}$-simultaneity is not.
\item [(h)]For arbitrary $K'$ and $K''$, $x^{K'}(A),t^{K'}(A)$ can be
expressed by $x^{K''}(A),t^{K''}(A)$ through a suitable Galilean
transformation
\item [(i)]For arbitrary $K'$ and $K''$, $\widetilde{x}^{K'}(A),\widetilde{t}^{K'}(A)$
can be expressed by $\widetilde{x}^{K''}(A),\widetilde{t}^{K''}(A)$
through a suitable Lorentz transformation.

\begin{lyxlist}{00.00.0000}
\item [$\vdots$]~
\end{lyxlist}
\end{lyxlist}
Moreover, they agree in the following observation (Relativity Principle): 

\begin{lyxlist}{0.0.0}
\item [(j)]The behaviour of similar systems co-moving as a whole with different
inertial frames, expressed in terms of the results of measurements
obtainable by means of co-moving measuring-rods and clocks (that is,
in terms of quantities $\widetilde{x}$ and $\widetilde{t}$) is the
same in every inertial frame of reference.
\end{lyxlist}
Combining this with (i),

\begin{lyxlist}{0.0.0}
\item [(k)]The laws of physics, expressed in terms of $\widetilde{x}$
and $\widetilde{t}$, must be given by means of Lorentz covariant
equations. 
\end{lyxlist}
Finally, they agree that 

\begin{lyxlist}{0.0.0}
\item [(l)]All facts about $\widetilde{x}$ and $\widetilde{t}$ (and,
consequently, all facts about $x$ and $t$) can be derived \emph{backward}
from (e) and (j).
\end{lyxlist}
To sum up symbolically, Lorentz theory and and special relativity
theory have identical assertions about both $\mathcal{M}$ and $\widetilde{\mathcal{M}}$:
they unanimously claim that $G_{1}\left(\mathcal{M}\right)\& G_{2}\left(\widetilde{\mathcal{M}}\right)$.

Finally, note that in an arbitrary inertial frame $K'$ for every
event $A$ the tags $x_{1}^{K'}(A)$, $x_{2}^{K'}(A),$ $x_{3}^{K'}(A)$,
$t^{K'}(A)$ can be expressed in terms of $\widetilde{x}_{1}^{K'}(A)$,
$\widetilde{x}_{2}^{K'}(A),$ $\widetilde{x}_{3}^{K'}(A)$, $\widetilde{t}^{K'}(A)$
and vice versa. Consequently, we can express the laws of physics---as
is done in special relativity---equally well in terms of the variables
$\widetilde{x}_{1},\widetilde{x}_{2},\widetilde{x}_{3},\widetilde{t}$
instead of the space and time tags $x_{1},x_{2},x_{3},t$. On the
other hand, we should emphasise that the one-to-one correspondence
between $\widetilde{x}_{1},\widetilde{x}_{2},\widetilde{x}_{3},\widetilde{t}$
and $x_{1},x_{2},x_{3},t$ also entails that the laws of physics (so
called {}``relativistic'' laws included) can be equally well expressed
in terms of the (traditional) space and time tags $x_{1},x_{2},x_{3},t$
\emph{}instead of the variables \emph{}$\widetilde{x}_{1},\widetilde{x}_{2},\widetilde{x}_{3},\widetilde{t}$.
In brief, \emph{physics could manage equally well with the classical
Galileo-invariant conceptions of space and time}.

\section{Comments}

\subsection{Are relativistic deformations real physical changes?}

Many believe that it is an essential difference between the two theories
that relativistic deformations like the Lorentz--FitzGerald contraction
and the time dilatation are real physical changes in Lorentz theory,
but there are no similar physical effects in special relativity. Let
us examine two typical argumentations.

According to the first argument the {}``Lorentz contraction/dilatation''
of a rod cannot be an objective physical deformation in relativity
theory, because it is a frame-dependent fact whether {}``the rod
is shrinking or expanding''. Consider a rod accelerated from the
sate of rest in reference frame $K'$ to the state of rest in reference
frame $K''$. According to relativity theory, {}``the rod shrinks
in frame $K'$ and, at the same time, expands in frame $K''$''.
But this is a contradiction, the argument says, if the deformation
was a real physical change. (In contrast, the argument says, Lorentz's
theory claims that {}``the length of a rod'' is a frame-independent
concept. Consequently, in Lorentz's theory, {}``the contraction/dilatation
of a rod'' can indeed be an objective physical change.) 

However, we have already clarified, that the terms {}``distance''
and {}``time'' have different meanings in relativity theory and
Lorentz's theory. Due to the difference between length and $\widetilde{\textrm{length}}$,
we must also differentiate dilatation from $\widetilde{\textrm{dilatation}}$,
contraction from $\widetilde{\textrm{contraction}}$, and so on. For
example, consider the reference frame of the \emph{etalons} $K$ and
another frame $K'$ moving relative to $K$. The following statements
are true about the {}``length'' of a rod accelerated from the sate
of rest in reference frame $K$ ($state_{1}$) to the state of rest
in reference frame $K'$ ($state_{2}$):\begin{eqnarray}
l^{K}\left(state_{1}\right) & > & l^{K}\left(state_{2}\right)\,\,\,\,\,\,\,\,\,\,\textrm{contraction in }K\label{eq:mindigaz1}\\
l^{K'}\left(state_{1}\right) & > & l^{K'}\left(state_{2}\right)\,\,\,\,\,\,\,\,\,\,\textrm{contraction in }K'\label{eq:mindigaz2}\\
\tilde{l}^{K}\left(state_{1}\right) & > & \tilde{l}^{K}\left(state_{2}\right)\,\,\,\,\,\,\,\,\,\,\widetilde{\textrm{contraction}}\textrm{ in }K\label{eq:mindigaz3}\\
\tilde{l}^{K'}\left(state_{1}\right) & < & \tilde{l}^{K'}\left(state_{2}\right)\,\,\,\,\,\,\,\,\,\,\widetilde{\textrm{dilatation}}\textrm{ in }K'\label{eq:mindigaz4}\end{eqnarray}
And there is no difference between relativity theory and Lorentz's
theory: \emph{all} of the four statements (\ref{eq:mindigaz1})--(\ref{eq:mindigaz4})
are true \emph{in both theories}. If, in Lorentz's theory, facts (\ref{eq:mindigaz1})--(\ref{eq:mindigaz2})
provide enough reason to say that there is a real physical change,
then the same facts provide enough reason to say the same thing in
relativity theory. And vice versa, if (\ref{eq:mindigaz3})--(\ref{eq:mindigaz4})
contradicted to the existence of real physical change of the rod in
relativity theory, then the same holds for Lorentz's theory. 

\begin{figure}
\begin{center}\includegraphics[%
  width=0.60\columnwidth]{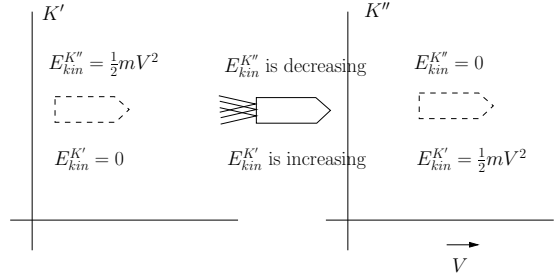}\end{center}

\caption{One and the same objective physical process is traced in the increase
of kinetic energy of the spaceship relative to frame $K'$, while
it is traced in the decrease of kinetic energy relative to frame $K''$\label{cap:forgas}}
\end{figure}

It should be mentioned, however, that there is no contradiction between
(\ref{eq:mindigaz3})--(\ref{eq:mindigaz4}) and the existence of
real physical change of the rod. Relativity theory and Lorentz's theory
unanimously claim that $\widetilde{\textrm{length}}$ is a relative
physical quantity. It is entirely possible that one and the same objective
physical change is traced in the increase of the value of a relative
quantity relative to one reference frame, while it is traced in the
decrease of the same quantity relative to another reference frame
(Fig~\ref{cap:forgas}). %
 (What is more, both, the value relative to one frame and the value
relative to the other frame, reflect objective features of the objective
physical process in question.)

According to the other wide-spread argument the relativistic deformations
cannot be real physical effects since they can be observed by an observer
also if the object is at rest but the observer is in motion at constant
velocity. And these {}``relativistic deformations'' cannot be explained
as real physical deformations of the object at rest---the argument
says. 

There is, however, a triple misunderstanding behind such an argument: 

\begin{itemize}
\item Of course, no real distortion is suffered by an object which is continuously
at rest relative to a reference frame $K'$, and, consequently, which
is continuously in motion at a constant velocity relative to another
frame $K''$. None of the observers can observe such a distortion.
For example, \begin{eqnarray*}
\widetilde{l}^{K'}\left(\textrm{distortion free rod at $\tilde{t}_{1}$}\right) & = & \widetilde{l}^{K'}\left(\textrm{distortion free rod at $\tilde{t}_{2}$}\right)\\
\widetilde{l}^{K''}\left(\textrm{distortion free rod at $\tilde{t}_{1}$}\right) & = & \widetilde{l}^{K''}\left(\textrm{distortion free rod at $\tilde{t}_{2}$}\right)\end{eqnarray*}
 
\item It is surely true,\begin{equation}
\widetilde{l}^{K'}\left(\textrm{distortion free rod}\right)\neq\widetilde{l}^{K''}\left(\textrm{distortion free rod}\right)\label{eq:azonosrud-masnezo}\end{equation}
 This fact, however, does not express a $\widetilde{\textrm{contraction}}$
of the rod---neither a real nor an apparent $\widetilde{\textrm{contraction}}$. 
\item On the other hand, inequality (\ref{eq:azonosrud-masnezo}) is a \emph{consequence}
of the real physical distortions suffered by the measuring equipments---with
which the $\widetilde{\textrm{space}}$ and $\widetilde{\textrm{time}}$
tags are operationally defined---when they are transfered from the
BIPM to the other reference frame in question.%
\footnote{For further details of what a moving observer can observe by means
of his or her distorted measuring equipments, see Bell 1983, pp. 75--76.%
}
\end{itemize}
Thus, relativistic deformations are real physical deformations also
in special relativity theory. One has to emphasise this fact because
it is an important part of the physical content of relativity theory.
It must be clear, however, that this conclusion is independent of
our main concern. What is important is the following: Lorentz's theory
and special relativity have identical assertions about length and
$\widetilde{\textrm{length}}$, duration and $\widetilde{\textrm{duration}}$,
shrinking and $\widetilde{\textrm{shrinking}}$, etc. Consequently,
whether or not these facts provide enough reason to say that the deformations
are real physical changes, the conclusion is common to both theories.

\subsection{The intuition behind the definitions\label{sub:The-intuition-behind}}

Different intuitions are behind the classical and the relativistic
definitions. As we have seen, both Lorentz theory and special relativity
{}``know'' about the distortions of measuring-rods and clocks when
they are transfered from the BIPM to the moving (relative to the BIPM)
reference frame $K'$. As it follows from the {}``compensatory view''
of the Lorentz theory and from the whole tradition of classical physics---according
to which if we are aware of some distortions of our measuring equipments,
we must take them into account and make corrections---the classical
definition (D5) takes into account these distortions. That is why
the space and time tags in $K'$ are defined through the original
space and time data, measured by the original distortion free measuring-rod
and clock, which are at rest relative to the BIPM. From the calculations
we made in the previous sections, it is easy to see, that we would
find the same $x^{K'}(A)$ and $t^{K'}(A)$ if we measured the space
and time tags with the co-moving measuring-rod and clock---obtaining
$\widetilde{x}^{K'}(A),\widetilde{t}^{K'}(A)$---and if we then made
corrections according to the known distortions of the equipments.

In contrast, in (D6) and (D7), relativity theory---although it is
aware of these deformations---ignores the distortions of measuring
rods and clocks, and defines the {}``space'' and {}``time'' tags
as they are measured by means of the distorted equipments. The physical
deformations in question are governed by some general physical laws
which apply to both the measuring-equipment and the object to be measured.
Therefore, it is no surprise that the {}``length'' of a moving,
consequently distorted, rod measured by co-moving, consequently distorted,
measuring-rod and clock is the same as the length of the corresponding
stationary rod measured with stationary measuring-rod and clock. The
{}``duration'' of a slowed down process in a moving object measured
with a co-moving, consequently slowed down, clock will be the same
as the duration of the same process in a similar object at rest, measured
with the original distortion free clock at rest. These and similar
observations lead us to believe in the relativity principle. 

These two distinct ways of thinking about {}``space'' and {}``time''
tags are clearly follow from Einstein's own words:

\begin{quote}
Let there be given a stationary rigid rod; and let its length be $l$
as measured by a measuring-rod which is also stationary. We now imagine
the axis of the rod lying along the axis of $x$ of the stationary
system of co-ordinates, and that a uniform motion of parallel translation
at velocity $v$ along the axis of $x$ in the direction of increasing
$x$ is then imparted to the rod. We now inquire as to the length
of the moving rod, and imagine its length to be ascertained by the
following two operations: 
\begin{quote}
(a) The observer moves together with the given measuring-rod and the
rod to be measured, and measures the length of the rod directly by
superposing the measuring-rod, in just the same way as if all three
were at rest. 

(b) By means of stationary clocks set up in the stationary system
and synchronising in accordance with §1, the observer ascertains at
what points of the stationary system the two ends of the rod to be
measured are located at a definite time. The distance between these
two points, measured by the measuring-rod already employed, which
in this case is at rest, is also a length which may be designated
{}``the length of the rod''. 
\end{quote}
In accordance with the principle of relativity the length to be discovered
by the operation (a) --- we will call it {}``the length of the rod
in the moving system'' --- must be equal to the length $l$ of the
stationary rod.

The length to be discovered by the operation (b) we will call {}``the
length of the (moving) rod in the stationary system''. This we shall
determine on the basis of our two principles, and we shall find that
it differs from $l$. (Einstein 1905)

\end{quote}
However, all these remarks on the intuitions behind definitions (D1)--(D8)
are not so important from the point of view of our main concern. What
is important is the simple fact that classical physics and relativity
theory use the terms {}``space'' and {}``time'' differently and
we have to make careful distinction whether we are about $\widetilde{x},\widetilde{t}$
or $x,t$.

\subsection{On the null result of the Michelson--Morley experiment}

It is clear from this careful distinction between $\widetilde{x},\widetilde{t}$
and $x,t$ that the null result of the Michelson--Morley experiment
simultaneously confirms \emph{both}, the classical rules of Galilean
kinematics for $x$ and $t$, and the violation of these rules for
the $\widetilde{\textrm{space}}$ and $\widetilde{\textrm{time}}$
tags $\widetilde{x},\widetilde{t}$. It confirms the classical addition
rule of velocities, on the one hand, and, on the other hand, it also
confirms that $\widetilde{\textrm{velocity}}$ of light is the same
in all frames of reference. Consider the following passage from Einstein:

\begin{quote}
A ray of light requires a perfectly definite time $T$ to pass from
one mirror to the other and back again, if the whole system be at
rest with respect to the aether. It is found by calculation, however,
that a slightly different time $T^{1}$ is required for this process,
if the body, together with the mirrors, be moving relatively to the
aether. And yet another point: it is shown by calculation that for
a given velocity $v$ with reference to the aether, this time $T^{1}$
is different when the body is moving perpendicularly to the planes
of the mirrors from that resulting when the motion is parallel to
these planes. Although the estimated difference between these two
times is exceedingly small, Michelson and Morley performed an experiment
involving interference in which this difference should have been clearly
detectable. But the experiment gave a negative result --- a fact very
perplexing to physicists. (Einstein 1920, p. 49)
\end{quote}
The {}``calculation'' that Einstein refers to is based on the Galilean
kinematics, that is, on the invariance of time and simultaneity, on
the invariance of distance, on the classical addition rule of velocities,
etc. The negative result was {}``very perplexing to physicists''
because their expectations were based on traditional concepts of space
and time, and they could not imagine other that if the speed of light
is $c$ relative to one inertial frame then the speed of the same
light signal cannot be the same $c$ relative to another reference
frame. So, it is obvious that {}``distance'', {}``time'' and {}``velocity''
in the above passage means the classical distance, time and velocity
defined in (D1), (D2) and (D5). 

On the other hand, Einstein continues this passage in the following
way:

\begin{quote}
Lorentz and FitzGerald rescued the theory from this difficulty by
assuming that the motion of the body relative to the aether produces
a contraction of the body in the direction of motion, the amount of
contraction being just sufficient to compensate for the difference
in time mentioned above. Comparison with the discussion in Section
11 shows that also from the standpoint of the theory of relativity
this solution of the difficulty was the right one. (Einstein 1920,
p. 49)
\end{quote}
What {}``rescued'' means here is that Lorentz and FitzGerald proved
that if the assumed deformations of moving bodies exist then the expected
result of the Michelson--Morley experiment is the null effect. On
the other hand, we have already clarified what Einstein also confirms
in the above quoted passage, that these deformations also derive from
the two basic postulates of special relativity. We can put these facts
together in the following schema: 

\[
\left.\begin{array}{c}
\underbrace{\begin{array}{ccc}
\left[\textrm{deformations}\right] & \& & \left[\begin{array}{c}
\textrm{Galilean kinematics}\\
\textrm{for }x,t\textrm{ (the speed}\\
\textrm{of light is NOT}\\
\textrm{the same in all}\\
\textrm{inertial frame}\end{array}\right]\end{array}}\\
\\\Updownarrow\\
\\\overbrace{\begin{array}{ccc}
\left[\begin{array}{c}
\textrm{Lorentz }\widetilde{\textrm{kinematics}}\\
\textrm{for }\tilde{x},\tilde{t}\textrm{ (the }\widetilde{\textrm{speed}}\\
\textrm{of light IS the same}\\
\textrm{in all inertial frame)}\end{array}\right] & \Rightarrow & \left[\textrm{deformations}\right]\end{array}}\end{array}\right\} \Rightarrow\left[\begin{array}{c}
\textrm{the result of the}\\
\textrm{Michelson-Morley}\\
\textrm{experiment must}\\
\textrm{be the null effect}\end{array}\right]\]

It is no surprise that the deformations can be {}``derived'' from
the Lorentz $\widetilde{\textrm{kinematics}}$. The \emph{physical}
information about the deformations suffered by objects accelerated
from one state of motion to another, say from the state of rest relative
to $K'$ to the state of rest relative to $K''$, is inbuilt into
the relationship between the tags $\tilde{x}^{K'}(A),\tilde{t}^{K'}(A)$
and $\tilde{x}^{K''}(A),\tilde{t}^{K''}(A)$. For these relations
are determined by the physical behaviour of measuring rods and clocks
during the acceleration and relaxation process. As Einstein warns
us, the Lorentz transformations, relating the $\widetilde{\textrm{space}}$
and $\widetilde{\textrm{time}}$ tags in different reference frames,
are nothing but physical laws governing the \emph{physical behaviour}
of the measuring-rods and clocks:

\begin{quote}
A Priori it is quite clear that we must be able to learn something
about the physical behaviour of measuring-rods and clocks from the
equations of transformation, for the magnitudes $z,y,x,t$ are nothing
more nor less than the results of measurements obtainable by means
of measuring-rods and clocks. (Einstein 1920, p. 35) 
\end{quote}

\subsection{The physical laws governing the relativistic deformations\label{sub:deformaciosszabalyok}}

FitzGerald, Lorentz%
\footnote{FitzGerald and Lorentz also made an attempt to understand how these
deformations actually come about from the molecular forces.%
} and Poincaré derived these laws from the requirement that the deformations
must explain the null result of the Michelson--Morley experiment.
Finally they arrived to the conclusion that the standard clock slows
down by factor $\sqrt{1-\frac{v^{2}}{c^{2}}}$ and that a rigid rod
suffers a contraction by factor $\sqrt{1-\frac{v^{2}}{c^{2}}}$ when
they are gently accelerated from $K$ to $K'$. As we have shown in
section~\ref{sec:Lorentz-theory-and}, this claim is equivalent with
the assertion that the $\widetilde{\textrm{space}}$ and $\widetilde{\textrm{time}}$
tags $\tilde{x}^{K''}(A),\tilde{t}^{K''}(A)$ measured by the co-moving
distorted equipments can be expressed from the similar tags $\tilde{x}^{K'}(A),\tilde{t}^{K'}(A)$
by a suitable Lorentz transformation.

Einstein derived the same transformation law from the assumption that
relativity principle generally holds%
\footnote{Whether or not relativity principle generally holds in relativistic
physics is a more complex question. See Szabó 2004.%
} and (or consequently) the $\widetilde{\textrm{velocity}}$ of a light
signal is the same in all inertial reference frames. 

These historic differences are not important from the point of view
of our main concern. What is important is that in both ways one can
derive exactly the same laws of deformations, exactly the same rules
for $x$ and $t$, and exactly the same rules for $\tilde{x}$ and
$\tilde{t}$.

\subsection{The conventionalist approach\label{sub:The-conventionalist-approach}}

According to the conventionalist thesis,%
\footnote{See Friedman 1983, p. 293; Einstein 1983, p. 35.%
} Lorentz's theory and Einstein's special relativity are two alternative
scientific theories which are equivalent on empirical level. Due to
the empirical underdeterminacy, the choice between these alternative
theories is based on external aspects.%
\footnote{Cf. Zahar 1973; Grünbaum 1974; Friedman 1983; Brush 1999; Janssen
2002.%
} Following Poincaré's similar argument about the relationship between
geometry, physics, and the empirical facts, the conventionalist thesis
asserts the following relationship between Lorentz theory and special
relativity:\begin{equation}
\begin{array}{ccccc}
\left[\begin{array}{c}
\textrm{classical}\\
E^{3}\times E^{1}\textrm{-theory}\\
\textrm{of space-time }\end{array}\right] & + & \left[\begin{array}{c}
\textrm{physical content}\\
\textrm{of Lorentz theory}\end{array}\right] & = & \left[\begin{array}{c}
\textrm{empirical}\\
\textrm{facts}\end{array}\right]\\
\\\left[\begin{array}{c}
\textrm{relativistic}\\
M^{4}\textrm{-theory}\\
\textrm{of space-time }\end{array}\right] & + & \left[\begin{array}{c}
\textrm{special relativistic}\\
\textrm{physics}\end{array}\right] & = & \left[\begin{array}{c}
\textrm{empirical}\\
\textrm{facts}\end{array}\right]\end{array}\label{eq:Poincare1}\end{equation}
Continuing the symbolic notations we used in the Introduction, denote
$\mathcal{Z}$ those objective features of physical reality that are
described by the alternative physical theories $P_{1}$ and $P_{2}$
in question. With these notations, the logical schema of the conventionalist
thesis can be described in the following way: We cannot distinguish
by means of the available experiments whether $G_{1}\left(\mathcal{M}\right)\& P_{1}\left(\mathcal{Z}\right)$
is true about the objective features of physical reality $\mathcal{M}\cup\mathcal{Z}$,
or $G_{2}\left(\mathcal{M}\right)\& P_{2}\left(\mathcal{Z}\right)$
is true about the \emph{same} objective features $\mathcal{M}\cup\mathcal{Z}$.
Schematically,\[
\begin{array}{rcl}
\left[G_{1}\left(\mathcal{M}\right)\right]+\left[P_{1}\left(\mathcal{Z}\right)\right] & = & \left[\begin{array}{c}
\textrm{empirical}\\
\textrm{facts}\end{array}\right]\\
\\\left[G_{2}\left(\mathcal{M}\right)\right]+\left[P_{2}\left(\mathcal{Z}\right)\right] & = & \left[\begin{array}{c}
\textrm{empirical}\\
\textrm{facts}\end{array}\right]\end{array}\]

However, it is clear from the previous sections that the terms {}``space''
and {}``time'' have different meanings in the two theories. Lorentz
theory claims $G_{1}\left(\mathcal{M}\right)$ about $\mathcal{M}$
and relativity theory claims $G_{2}\left(\widetilde{\mathcal{M}}\right)$
about some other features of reality $\widetilde{\mathcal{M}}$. Of
course, this terminological confusion also appears in the physical
assertions. Let us symbolise with $\mathcal{Z}$ the objective features
of physical reality, such as the length of a rod, etc., described
by physical theory $P_{1}$. And let $\widetilde{\mathcal{Z}}$ denote
some (partly) different features of reality described by $P_{2}$,
such as the $\widetilde{\textrm{length}}$ of a rod, etc. Now, as
we have seen, both theories actually claim that $G_{1}\left(\mathcal{M}\right)\& G_{2}\left(\widetilde{\mathcal{M}}\right)$.
It is also clear that, for example, within Lorentz's theory, we can
legitimately query the $\widetilde{\textrm{length}}$ of a rod. For
Lorentz's theory has complete description of the behaviour of a moving
rigid rod, as well as the behaviour of a moving clock and measuring-rod.
Therefore, it is no problem in Lorentz's theroy to predict the result
of a measurement of the {}``length'' of the rod, if the measurement
is performed with a co-moving measuring equipments, according to empirical
definition (D7). This prediction will be axactly the same as the prediction
of special relativity. And vice versa, special relativity would have
the same prediction for the length of the rod as the prediction of
the Lorentz theory. That is to say, the physical contents of Lorentz's
theory and special relativity also are identical: both claim that
$P_{1}\left(\mathcal{Z}\right)\& P_{2}\left(\widetilde{\mathcal{Z}}\right)$.
So we have the following:\[
\begin{array}{rcl}
\left[G_{1}\left(\mathcal{M}\right)\& G_{2}\left(\widetilde{\mathcal{M}}\right)\right]+\left[P_{1}\left(\mathcal{Z}\right)\& P_{2}\left(\widetilde{\mathcal{Z}}\right)\right] & = & \left[\begin{array}{c}
\textrm{empirical}\\
\textrm{facts}\end{array}\right]\\
\\\left[G_{1}\left(\mathcal{M}\right)\& G_{2}\left(\widetilde{\mathcal{M}}\right)\right]+\left[P_{1}\left(\mathcal{Z}\right)\& P_{2}\left(\widetilde{\mathcal{Z}}\right)\right] & = & \left[\begin{array}{c}
\textrm{empirical}\\
\textrm{facts}\end{array}\right]\end{array}\]
In other words, since there are no two different theories, there is
\emph{no choice}, based neither on internal nor on external aspects.

\subsection{Methodological remarks}

\begin{enumerate}
\item It worth while emphasising that my argument is based on the following
very weak {}``operationalist'' premise: physical terms, assigned
to measurable physical quantities, have different meanings if they
have different empirical definitions. This premise is one of the fundamental
pre-assumptions of Einstein's 1905 paper (see, for example, the quotation
in section~\ref{sub:The-intuition-behind}) and is widely accepted
among physicists. Without clear empirical definition of the measurable
physical quantities a physical theory cannot be empirically confirmable
or disconfirmable. In itself, this premise is not yet equivalent to
operationalism or verificationalism. It does not generally imply that
a statement is necessarily meaningless if it is neither analytic nor
empirically verifiable. However, when the physicist assigns time and
space tags to an event, relative to a reference frame, (s)he is already
after all kinds of metaphysical considerations about {}``What is
space and what is time?'' and means definite physical quantities
with already settled empirical meanings.
\item In saying that the meanings of the words {}``space'' and {}``time''
are different in relativity theory and in classical physics, it is
necessary to be careful of a possible misunderstanding. I am talking
about something entirely different from the incommensurability thesis
of the relativist philosophy of science.%
\footnote{See Kuhn 1970, Chapter X; Feyerabend 1970.%
} How is it that relativity makes any assertion about classical space
and time, and vice versa, how can Lorentz's theory make assertions
about quantities which are not even defined in the theory? As we have
seen, each of the two theories is sufficiently complete account of
physical reality to make predictions about those features of reality
that correspond---according to the empirical definitions---to the
variables used by the other theory, and we can \emph{compare} these
predictions. For example, within Lorentz's theory, we can legitimately
query the reading of a clock slowly transported in $K'$ from one
place to another. That exactly is what we calculated in section~\ref{sec:Lorentz-theory-and}.
Similarly, in relativity theory, we can legitimately query the $\widetilde{\textrm{space}}$
and $\widetilde{\textrm{time}}$ tags of an event in the reference
frame of the \emph{etalons} and the result of a Galilean transformation.
This is a fair calculation, in spite of the fact that the result so
obtained is not explicitly mentioned and named in the theory. This
is what we actually did. And the conclusion was that not only are
the two theories commensurable, but they provide completely identical
accounts for space and time.
\end{enumerate}

\subsection{Privileged reference frame}

Due to the popular/textbook literature on relativity theory, there
is a widespread aversion to a privileged reference frame. However,
like it or not, there is a privileged reference frame in both special
relativity and classical physics. It is the frame of reference in
which the \emph{etalons} are at rest. This privileged reference frame,
however, has nothing to do with the concepts of {}``absolute rest''
or the aether, and it is not privileged by nature, but it is privileged
by the trivial semantical convention providing meanings for the terms
{}``distance'' and {}``time'', by the fact that of all possible
measuring-rod-like and clock-like objects floating in the universe,
we have chosen the ones floating together with the International Bureau
of Weights and Measures in Paris. In Bridgman's words:

\begin{quote}
It cannot be too strongly emphasised that there is no getting away
from preferred operations and unique standpoint in physics; the unique
physical operations in terms of which interval has its meaning afford
one example, and there are many others also. (Bridgman 1936, p. 83)
\end{quote}
Many believe that one can avoid a reference to the \emph{etalons}
sitting in a privileged reference frame by defining, for example,
the unit of $\widetilde{\textrm{time}}$ for an arbitrary (moving)
frame of reference $K'$ through a cesium clock, or the like, co-moving
with $K'$. In this way, one needs not to refer to a standard clock
accelerated from the reference frame of the \emph{etalons} into reference
frame $K'$. But further thought reveals that such a definition has
several difficulties. For if this operation is regarded as a convenient
way of measuring $\widetilde{\textrm{time}}$, then we still have
$\widetilde{\textrm{time}}$ in the theory, together with the privileged
reference frame of the \emph{etalons}. If, however, this operation
is regarded as the empirical definition of a physical quantity, then
it must be clear that this quantity is not $\widetilde{\textrm{time}}$
but a new physical quantity, say $\widetilde{\widetilde{\textrm{time}}}$.
In order to establish any relationship between $\widetilde{\widetilde{\textrm{time}}}$
tags belonging to different reference frames, it is a must to use
an {}``\emph{etalon} cesium clock'' as well as to refer to its behaviour
when accelerated from one inertial frame into the other.

\subsection{The physics of moving objects}

Although special relativity did not produce a new theory of space
and time, both special relativity and Lorentz theory enrich our knowledge
of the physical world with \emph{the physics of objects moving at
constant velocities}---in accordance with the title of Einstein's
original 1905 paper. The essential physical content of their discoveries
is that physical objects suffer distortions when they are accelerated
from one inertial frame to the other, and that these distortions satisfy
some uniform laws. From the simplest rules discussed in section~\ref{sub:deformaciosszabalyok},
Lorentz, Poincaré and Einstein concluded with the general validity
of the relativity principle. And the relativity principle together
with the Lorentz transformation of $\widetilde{\textrm{space}}$ and
$\widetilde{\textrm{time}}$ provide the general description of the
behavior of moving physical systems. In his 1905 paper, Einstein shows
examples of how to understand and how to apply these principles (see,
for example, the quotation in \ref{sub:The-intuition-behind}): Let
$\mathcal{E}'$ be a set of differential equations describing the
behaviour of the system in question in an arbitrary reference frame
$K'$. Let $\psi'_{0}$ denote a set of (initial) conditions, such
that the solution determined by $\psi'_{0}$ describes the behaviour
of the system when it is, as a whole, at rest relative to $K'$. Let
$\psi'_{\widetilde{v}}$ be a set of conditions which corresponds
to the solution describing the same system in uniform motion at velocity
$\widetilde{v}$ relative to $K'$. To be more exact, $\psi'_{\widetilde{v}}$
corresponds to a solution of $\mathcal{E}'$ that describes the same
behaviour of the system as $\psi'_{0}$ but in superposition with
a collective translation at velocity $\widetilde{v}$. Denote $\mathcal{E}''$
and $\psi''_{0}$ the equations and conditions obtained from $\mathcal{E}'$
and $\psi'_{0}$ by substituting every $\widetilde{x}^{K'}$ with
$\widetilde{x}^{K''}$ and $\widetilde{t}^{K'}$ with $\widetilde{t}^{K''}$.
Denote $\Lambda_{\widetilde{v}}\left(\mathcal{E}'\right),\Lambda_{\widetilde{v}}\left(\psi'_{\widetilde{v}}\right)$
the set of equations and conditions expressed in terms of the double-primed
variables, applying the Lorentz transformations. Now, what the (relativistic
version of) relativity principle states is that the laws of physics
describing the behaviour of moving objects are such that they satisfy
the following relationships: \begin{eqnarray}
\Lambda_{\widetilde{v}}\left(\mathcal{E}'\right) & = & \mathcal{E}''\label{eq:Lequivalent1}\\
\Lambda_{\widetilde{v}}\left(\psi'_{\widetilde{v}}\right) & = & \psi''_{0}\label{eq:Lequivalent2}\end{eqnarray}

To make more explicit how this principle provides a useful method
in the description of the deformations of physical systems when they
are accelerated from one inertial frame $K'$ into some other $K''$,
consider the following situation: Assume we know the relevant physical
equations and know the solution of the equations describing the physical
properties of the object in question when it is at rest in $K'$:
$\mathcal{E}',\psi'_{0}$. We now inquire as to the same description
of the object when it is moving at a given constant $\widetilde{\textrm{velocity}}$
relative to $K'$. If (\ref{eq:Lequivalent1})--(\ref{eq:Lequivalent2})
is true, then we can solve the problem in the following way. Simply
take $\mathcal{E}'',\psi''_{0}$---by putting one more prime on each
variable---and express $\psi'_{\widetilde{v}}$ from (\ref{eq:Lequivalent2})
by means of the inverse Lorentz transformation: $\psi'_{\widetilde{v}}=\Lambda_{\widetilde{v}}^{-1}\left(\psi''_{0}\right)$.%
\footnote{Actually, the situation is much more complex. Whether or not the solution
thus obtained is correct depends on the details of the relaxation
process after the acceleration of the system. (See Szabó 2004)%
} This is the way we usually solve problems such as the electromagnetic
field of a moving point charge, the Lorentz contraction of a rigid
body, the loss of phase suffered by a moving clock, the dilatation
of the mean life of a cosmic ray $\mu$-meson, etc.

\subsection{The aether}

Many of those who admit the {}``empirical equivalence'' of Lorentz's
theory and special relativity argue that the latter is {}``incomparably
more satisfactory'' because it has no reference to the aether (\emph{e.g.}
Einstein 1920, p. 50). As it is obvious from the previous sections,
we did not make any reference to the aether in the logical reconstruction
of Lorentz's theory. It is however a historic fact that Lorentz did.
In this section, I want to clarify that the concept of aether is merely
a verbal decoration in Lorentz theory, which can be interesting for
the historians, but negligible  from the point of view of recent logical
reconstructions. 

One can find various verbal formulations of the relativity principle
and Lorentz-covariance. In order to compare these formulations, let
us introduce the following notations:

\begin{lyxlist}{00.00.00000000}
\item [$A\left(K',K''\right):=$]The laws of physics in inertial frame
$K'$ are such that the laws describing a physical system co-moving
with frame $K''$ are obtainable by solving the problem for the similar
physical system at rest relative to $K'$ and perform the following
substitutions:\begin{eqnarray}
\widetilde{x}_{1}^{K'} & \mapsto & \alpha_{1}=\widetilde{x}_{1}^{K'}\nonumber \\
\widetilde{x}_{2}^{K'} & \mapsto & \alpha_{2}=\widetilde{x}_{2}^{K'}\nonumber \\
\widetilde{x}_{3}^{K'} & \mapsto & \alpha_{3}=\frac{\widetilde{x}_{3}^{K'}-\widetilde{v}\widetilde{t}^{K'}}{\sqrt{1-\frac{\widetilde{v}^{2}}{c^{2}}}}\label{eq_helyettesites}\\
\widetilde{t}^{K'} & \mapsto & \tau=\frac{\widetilde{t}^{K'}-\frac{\widetilde{v}}{c^{2}}\widetilde{x}_{3}^{K'}}{\sqrt{1-\frac{\widetilde{v}^{2}}{c^{2}}}}\nonumber \end{eqnarray}

\item [$B\left(K',K''\right):=$]The laws of physics in $K'$ are such
that the mathematically introduced variables $\alpha_{1},\alpha_{2},\alpha_{3},\tau$
in (\ref{eq_helyettesites}) are equal to $\widetilde{x}_{1}^{K''},\widetilde{x}_{2}^{K''},\widetilde{x}_{3}^{K''},\widetilde{t}^{K''}$,
that is, the {}``space'' and {}``time'' tags obtained by means
of measurements in $K''$, performed with the same measuring-rods
and clocks we used in $K'$ after that they were transfered from $K'$
into $K''$, ignoring the fact that the equipments undergo deformations
during the transmission.
\item [$C\left(K',K''\right):=$]The laws of physics in $K'$ are such
that the laws of physics empirically ascertained by an observer in
$K''$, describing the behaviour of physical objects co-moving with
$K''$, expressed in variables $\widetilde{x}_{1}^{K''},\widetilde{x}_{2}^{K''},\widetilde{x}_{3}^{K''},\widetilde{t}^{K''}$,
have the same forms as the similar empirically ascertained laws of
physics in in $K'$, describing the similar physical objects co-moving
with $K'$, expressed in variables $\widetilde{x}_{1}^{K'},\widetilde{x}_{2}^{K'},\widetilde{x}_{3}^{K'},\widetilde{t}^{K'}$,
if the observer in $K''$ performs the same measurement operations
as the observer in $K'$ with the same measuring equipments transfered
from $K'$ to $K''$, ignoring the fact that the equipments undergo
deformations during the transmission.
\end{lyxlist}
It is obvious that \[
A\left(K',K''\right)\,\&\, B\left(K',K''\right)\Rightarrow C\left(K',K''\right)\]
 So, let us restrict our considerations on the more fundamental \[
A\left(K',K''\right)\,\&\, B\left(K',K''\right)\]
 Taking this statement, the usual Einsteinian formulation of the relativity
principle is the following:

\[
\textrm{Einstein's Relativity Principle}=\left(\forall K'\right)\left(\forall K''\right)\left[A\left(K',K''\right)\,\&\, B\left(K',K''\right)\right]\]

Many believe that this version of relativity principle is essentially
different from the similar principle of Lorentz, since Lorentz's principle
makes explicit reference to the motion relative to the aether. Using
the above introduced notations, it says the following:

\[
\textrm{Lorentz's Principle}=\left(\forall K''\right)\left[A\left(\textrm{aether},K''\right)\,\&\, B\left(\textrm{aether},K''\right)\right]\]
It must be clearly seen, however, that Lorentz's aether hypothesis
is logically independent from the actual physical content of his theory.
In fact, as a little reflection reveals, \emph{Lorentz's principle
and Einstein's relativity principle are logically equivalent to each
other.} It is trivially true that \begin{eqnarray*}
\textrm{Einstein's Relativity Principle} & = & \left(\forall K'\right)\left(\forall K''\right)\left[A\left(K',K''\right)\,\&\, B\left(K',K''\right)\right]\\
 & \Rightarrow & \left(\forall K''\right)\left[A\left(\textrm{aether},K''\right)\,\&\, B\left(\textrm{aether},K''\right)\right]\\
 & = & \textrm{Lorentz's Principle}\end{eqnarray*}
It follows from the meaning of $A\left(K',K''\right)$ and $B\left(K',K''\right)$
that\begin{eqnarray*}
 &  & \left(\exists K'\right)\left(\forall K''\right)\left[A\left(K',K''\right)\,\&\, B\left(K',K''\right)\right]\\
 & \Rightarrow & \left(\forall K'\right)\left(\forall K''\right)\left[A\left(K',K''\right)\,\&\, B\left(K',K''\right)\right]\end{eqnarray*}
Consequently, \begin{eqnarray*}
\textrm{Lorentz's Principle} & = & \left(\forall K''\right)\left[A\left(\textrm{aether},K''\right)\,\&\, B\left(\textrm{aether},K''\right)\right]\\
 & \Rightarrow & \left(\exists K'\right)\left(\forall K''\right)\left[A\left(K',K''\right)\,\&\, B\left(K',K''\right)\right]\\
 & \Rightarrow & \left(\forall K'\right)\left(\forall K''\right)\left[A\left(K',K''\right)\,\&\, B\left(K',K''\right)\right]\\
 & = & \textrm{Einstein's Relativity Principle}\end{eqnarray*}
Thus, it is Lorentz's principle itself---the verbal formulation of
which refers to the aether---that renders any claim about the aether
a logically separated hypothesis outside of the scope of the factual
content of both Lorentz theory and special relativity. The role of
the aether could be played by anything else. As both theories claim,
it follows from the empirically confirmed laws of physics that physical
systems undergo deformations when they are transferred from one inertial
frame $K'$ to another frame $K''$. One could say, these deformations
are caused by the transmission of the system from $K'$ to $K''$.
You could say they are caused by the {}``wind of aether''. By the
same token you could say, however, that they are caused by {}``the
wind of \emph{anything}'', since if the physical system is transfered
from $K'$ to $K''$ then its state of motion changes relative to
an arbitrary third frame of reference.

On the other hand, it must be mentioned that special relativity does
not exclude the existence of the aether.%
\footnote{Not to mention that already in 1920 Einstein himself argues for the
existence of some kind of aether. (See Reignier 2000)%
} Neither does the Michelson--Morley experiment. If special relativity/Lorentz
theory is true then there must be no indication of the motion of the
interferometer relative to the aether. Consequently, the fact that
we do not observe this motion is not a challenge for the aether theorist.
Thus, the hypothesis about the existence of aether is logically independent
of both Lorentz theory and special relativity.

\subsection{Symmetry principle and heuristic value}

Finally, it worth while mentioning that Lorentz's theory and special
relativity, as completely identical theories, offer the same symmetry
principles and heuristic power. As we have seen, both theories claim
that quantities $\widetilde{x}^{K'},\widetilde{t}^{K'}$ in an arbitrary
$K'$ and the similar quantities $\widetilde{x}^{K''},\widetilde{t}^{K''}$
in another arbitrary $K''$ are related through a suitable Lorentz
transformation. This fact in conjunction with the relativity principle
implies that laws of physics are to be described by Lorentz covariant
equations, if they are expressed in terms of variables $\widetilde{x}$
and $\widetilde{t}$, that is, in terms of the results of measurements
obtainable by means of the corresponding co-moving equipments---which
are distorted relative to the \emph{etalons}. There is no difference
between the two theories that this $\widetilde{\textrm{space}}$-$\widetilde{\textrm{time}}$
symmetry provides a valuable heuristic aid in the search for new laws
of nature.

\section{Conclusion}

With these comments I have completed the argumentation for my basic
claim that special relativity and Lorentz theory are completely identical
in both senses, as theories about space-time and as theories about
the behaviour of moving physical objects. Consequently, in comparison
with the classical Galileo-invariant conceptions, special relativity
theory does not tell us anything new about space and time. As we have
seen, the longstanding belief that it does is the result of a simple
but subversive terminological confusion.

\subsection*{Acknowledgement}

The research was partly supported by the OTKA Foundation, No. T 037575
and No. T 032771. I am grateful to the Netherlands Institute for Advanced
Study (NIAS) for providing me with the opportunity, as a Fellow-in-Residence,
to complete this paper.

\section*{References}

\setlength{\itemindent}{-15pt}\setlength{\leftmargin}{15pt}

\begin{description}
\item [Bell,]J. S. (1987): \emph{Speakable and unspeakable in quantum mechanics},
Cambridge University Press, Cambridge.
\item [Bell,]J. S. (1992): George Francis FitzGerald, \emph{Physics World}
\textbf{5}, pp. 31-35.
\item [Bridgman,]P. (1927): \emph{The Logic of Modern Physics}, MacMillan,
New York.
\item [Brown,]H. R. and Pooley, O. (2001): The origin of space-time metric:
Bell's 'Lorentzian pedagogy' and its significance in general relativity,
in \emph{Physics meets philosophy at the Planck scale. Contemporary
theories in quantum gravity}, C. Calleander and N. Huggett (eds.),
Cambridge University Press, Cambridge.
\item [Brown,]H. R (2001): The origins of length contraction: I. The FitzGerald-Lorentz
deformation, http://philsci-archive.pitt.edu/archive/00000218.
\item [Brown,]H. R. (2003): Michelson, FitzGerald and Lorentz: the origins
of relativity revisited, http://philsci-archive.pitt.edu/archive/00000987.
\item [Brush,]S. G. (1999): Why was Relativity Accepted?, \emph{Physics
in Perspective} \textbf{1}, pp. 184--214.
\item [Einstein,]A (1905): Zur Elektrodynamik bewegter Körper, \emph{Annalen
der Physik} \textbf{17}, p. 891.
\item [Einstein,]A. (1920): \emph{Relativity: The Special and General Theory},
H. Holt and Company, New York.
\item [Einstein,]A. (1983): \emph{Sidelights on relativity}, Dover, New
York.
\item [Feyerabend,]P. K. (1970): Consolation for the Specialist, in \emph{Criticism
and the Growth of Knowledge}, I. Lakatos and A. Musgrave (eds.), Cambridge
University Press, Cambridge, pp. 197--230.
\item [Friedman,]M. (1983): \emph{Foundations of Space-Time Theories --
Relativistic Physics and Philosophy of Science}, Princeton University
Press, Princeton.
\item [Grünbaum,]A. (1974): \emph{Philosophical Problems of Space and Time},
Boston Studies in the Philosophy of Science, Vol. XII. (R. S. Cohen
and M. W. Wartofsky, eds.) D. Reidel, Dordrecht.
\item [Jánossy,]L. (1971): \emph{Theory of relativity based on physical
reality,} Akadémiai Kiadó, Budapest.
\item [Janssen,]M. (2002): Reconsidering a Scientific Revolution: The Case
of Einstein \emph{versus} Lorentz, \emph{Physics in Perspective} \textbf{4},
pp. 421--446
\item [Kuhn,]T. S. (1970): \emph{The Structure of Scientific Revolution},
University of Chicago Press, Chicago.
\item [Lorentz,]H. A. (1904): Electromagnetic phenomena in a system moving
with any velocity less than that of light, \emph{Proc. R. Acad. Amsterdam}
\textbf{6}, p. 809.
\item [Malament,]D. (1977): Causal Theories of Time and the Conventionality
of Simultaneity, \emph{No\^{u}s} \textbf{11}, p. 293.
\item [Reichenbach,]H. (1956): \emph{The Direction of Time}, University
of California Press, Berkeley.
\item [Reignier,]J. (2000): The birth of special relativity. {}``One more
essay on the subject'', arXiv:physics/0008229.
\item [Salmon,]W. C. (1977): The Philosophical Significance of the One-Way
Speed of Light, \emph{No\^{u}s} \textbf{11}, p. 253.
\item [Szabó,]L. E. (2004): On the meaning of Lorentz covariance, \emph{Foundations
of Physics Letters} \textbf{17}, p. 479.
\item [Tonnelat,]M. A. (1971): \emph{Histoire du principe de relativité},
Flammarion, Paris.
\item [Zahar,]E. (1973): Why did Einstein's Programme Supersede Lorentz's?,
\emph{British Journal for the Philosophy of Science}, \textbf{24}
pp. 95--123, 223--262.\end{description}

\end{document}